\documentclass[english,british]{jfm}
\usepackage[T1]{fontenc}
\usepackage[applemac]{inputenc}
\usepackage{color}
\usepackage{babel}
\usepackage{refstyle}
\usepackage{textcomp}
\usepackage{amstext}
\usepackage{wasysym}
\usepackage{graphicx}
\usepackage[unicode=true,pdfusetitle,
 bookmarks=true,bookmarksnumbered=false,bookmarksopen=false,
 breaklinks=false,pdfborder={0 0 1},backref=false,colorlinks=true]
 {hyperref}
\hypersetup{
 allcolors=blue}

\makeatletter

\AtBeginDocument{\providecommand\figref[1]{\ref{fig:#1}}}
\RS@ifundefined{subref}
  {\def\RSsubtxt{section~}\newref{sub}{name = \RSsubtxt}}
  {}
\RS@ifundefined{thmref}
  {\def\RSthmtxt{theorem~}\newref{thm}{name = \RSthmtxt}}
  {}
\RS@ifundefined{lemref}
  {\def\RSlemtxt{lemma~}\newref{lem}{name = \RSlemtxt}}
  {}

\usepackage{graphicx}
\usepackage{natbib}
\ifCUPmtlplainloaded \else
  \checkfont{eurm10}
  \iffontfound
    \IfFileExists{upmath.sty}
      {\typeout{^^JFound AMS Euler Roman fonts on the system,
                   using the 'upmath' package.^^J}%
       \usepackage{upmath}}
      {\typeout{^^JFound AMS Euler Roman fonts on the system, but you
                   dont seem to have the}%
       \typeout{'upmath' package installed. JFM.cls can take advantage
                 of these fonts,^^Jif you use 'upmath' package.^^J}%
       \providecommand\upi{\pi}%
      }
  \else
    \providecommand\upi{\pi}%
  \fi
\fi

\ifCUPmtlplainloaded \else
  \checkfont{msam10}
  \iffontfound
    \IfFileExists{amssymb.sty}
      {\typeout{^^JFound AMS Symbol fonts on the system, using the
                'amssymb' package.^^J}%
       \usepackage{amssymb}%

      }{}
  \fi
\fi

\ifCUPmtlplainloaded \else
  \IfFileExists{amsbsy.sty}
    {\typeout{^^JFound the 'amsbsy' package on the system, using it.^^J}%
     \usepackage{amsbsy}}
    {\providecommand\boldsymbol[1]{\mbox{\boldmath $##1$}}}
\fi

\providecommand\bnabla{\boldsymbol{\nabla}}

\newcommand\Real{\mbox{Re}} % cf plain TeX's \Re and Reynolds number
\newcommand\Imag{\mbox{Im}} % cf plain TeX's \Im
\newsavebox{\astrutbox}
\sbox{\astrutbox}{\rule[-5pt]{0pt}{20pt}}

\newcommand\eg{e.g.\ }

\newcommand\sech{\mathrm{sech}}
\usepackage{lmodern}

\makeatother

\begin{document}

\title[Velocity field in parametrically excited solitary waves]{Measurement of velocity field in\\
parametrically excited solitary waves}

\author[L. Gordillo and N. Mujica]{Leonardo Gordillo$^{1,2}$\thanks{Email address for correspondence: leonardo.gordillo@univ-paris-diderot.fr}
and Nicolás Mujica$^{1}$}

\affiliation{$^{1}$Departamento de Física, Facultad de Ciencias Físicas y Matemáticas,
Universidad de Chile, Casilla 487-3, Santiago, Chile\\[\affilskip]$^{2}$Laboratoire
\textquotedbl{}Matière et Systèmes Complexes\textquotedbl{} (MSC),
UMR 7057 CNRS, Université Paris 7 Diderot, 75205 Paris Cedex 13, France}

\date{?; revised ?; accepted ?. - To be entered by editorial office}

\pubyear{2014}

\volume{??}

\pagerange{??}
\maketitle
\begin{abstract}
Parametrically excited solitary waves emerge as localized structures
in high-aspect-ratio free surfaces subject to vertical vibrations.
Herein, we provide the first experimental characterization of the
hydrodynamics of these waves using Particle Image Velocimetry. We
show that the underlying velocity field of parametrically excited
solitary waves is mainly composed by an oscillatory velocity field.
Our results confirm the accuracy of Hamiltonian models with added
dissipation in describing this field. Remarkably, our measurements
also uncover the onset of a streaming velocity field which is shown
to be as important as other crucial nonlinear terms in the current
theory. The observed streaming pattern is particularly interesting
due to the presence of oscillatory meniscii.
\end{abstract}

\section{Introduction}

Parametric instabilities in spatially extended systems can generate
waves by their resonance with an external driving, which is a universal
mechanism to generate structures in dissipative systems. In hydrodynamics,
these structures satisfy a simple rule: energy losses due to viscous
effects are compensated by the external injection of energy, \emph{\eg}
by means of vertical vibrations. This balance can create extended
or solitary structures that remain stable as long as the system is
driven by the external force. In particular, solitary waves emerge
in high-aspect-ratio free surfaces subject to vertical vibrations
as a result of exciting the system at double the frequency of the
first transverse mode. They become stable only after perturbing the
free surface. Although these waves keep the sloshing motion features
of the first transverse mode, their motion is highly localized in
the longitudinal direction instead of involving the whole surface
\citep{1984PhRvL..52.1421W}. This solitary wave is usually referred
as non-propagating hydrodynamic soliton or parametrically excited
solitary waves. The spatial envelopes are steady, very stable and
do not propagate in contrast with classical hydrodynamic solitons.
Remarkably, this dynamic behaviour can be found in several hydrodynamic
systems: oscillons in Faraday configuration \citep{2000PhRvL..85..756A}
or solitary waves in vibrated Hele-Shaw cells \citep{2011PhRvL.107b4502R}
display spatial envelopes with similar spatiotemporal features.

Parametrically excited solitary waves are modelled in terms of the\emph{
parametric dissipative nonlinear Schrödinger equation} (pdNLSe), derived
by \citealp{1984JFM...148..451M}. This equation captures the minimum
requirements for parametrically sustained one-dimensional solitary
structures. Hence, its scope goes far beyond hydrodynamics \citep[see \emph{e.g.} ][]{1991EL.....15..113B,1992PhRvL..68.1730D}.
In this regard, recent studies have been focused on providing an exhaustive
analysis of the pdNLSe using mathematical and numerical techniques.
Experiments have also been used for this purpose \citep[see][]{Zhang:2007cv,2011EPJD...62...39G,2011PhRvE..84c6205C}.
On the contrary, some fundamental issues such as the validity of the
approximations that yield the pdNLSe have remained unaddressed, with
rare systematic comparisons between predictions and measurements \citep{1999PhLA..255..272C}.
For worse, all the experimental characterizations that can be found
in the literature have been achieved using a single technique, \emph{i.e.}
by tracking the free surface, a measurement useful for outlining solitary-waves
stability and interaction laws but blind to potential underlying phenomena.
This is a critical issue, since most steps involved in the pdNLSe
derivation lie on strong hypotheses from fluid dynamics. Uncovering
the velocity field beneath the structures is hence fundamental for
a comprehensive experimental analysis of parametrically excited solitary
waves.

In this article, we present experimental results concerning parametrically
excited solitary waves with a close view on the hydrodynamics. The
velocity fields that support the localized cross waves have been measured
using Particle Image Velocimetry (PIV). The article is organized
as follows. In \S\ref{sec:Theory}, we outline the pdNLSe theoretical
model. In \S\ref{sec:Setup}, we describe briefly the experimental
setup. A summary of our most significant experimental results can
be found in \S\ref{sec:Results}. Finally, discussion and conclusions,
including comparisons with pdNLSe theoretical predictions, are given
in \S\ref{sec:Discussion}.

\section{Theory\label{sec:Theory}}

Consider a fluid layer of depth $d$ in an infinite channel of breadth
$b$. The channel is oriented in the $x$ axis (walls at $y=\pm\frac{1}{2}b$
and $z=-d$) and forced to oscillate vertically at frequency $2\omega$
and acceleration amplitude $\Gamma_{0}$. The vertical acceleration
of the channel is accordingly $\Gamma\left(t\right)=-\Gamma_{0}\cos2\omega t$.
Let us assume that $\omega$ approaches $\omega_{01}$, the first-transverse-mode
frequency so the $\left(0,1\right)$ mode is parametrically excited.
The linear theory of gravity waves for inviscid flows provides a good
estimation for the $\left(0,1\right)$-mode frequency, $\omega_{01}=\sqrt{gk\tau}$,
where $g$ is the acceleration gravity, $k=\upi/b$ is the wavenumber
and $\tau\equiv\tanh kd.$ The parametric forcing can thus be characterized
in terms of two dimensionless parameters: the detuning $\nu=\frac{1}{2}\left(\omega^{2}/\omega_{01}^{2}-1\right)$
and the normalized acceleration amplitude $\gamma=\frac{1}{4}\Gamma_{0}/g$.

Free-surface waves on constant-depth inviscid flows are known to satisfy
Hamiltonian properties \citep[cf.][]{1977JFM....83..153M}. This can
be used as a point of departure for deriving amplitude equations in
such kind of system. However, realistic setups cannot be considered
as conservative: external forcing is required to create structures.
In any case, although viscous effect are neglected \emph{a priori}
in Hamiltonian formulations, adding linear dissipation in amplitude
equations seems to be enough for modelling slightly viscous flows
\citep{1976JFM....75..419M}. The reason is simple: in this kind of
flow, the motion is basically inviscid all over the space except in
thin boundary layers. Energy is thus dissipated without affecting
the general features of the waves \citep[cf.][]{Miles:1967vp}. For
instance, the decay rate for the $\left(0,1\right)$ mode, $\alpha_{01}$,
can be estimated from boundary layer analysis \citep{1984JFM...148..451M}.
In our problem, this provides an extra dimensionless parameter, the
damping rate $\mu=\alpha_{01}/\omega_{01}$. 

Using the stated hypotheses, it can be shown rigorously that the first
transverse mode in an infinite channel can be modelled with \citep{1984JFM...148..451M}
\begin{equation}
\mathrm{i}\left(\partial_{T}\psi+\mu\psi\right)=\nu\psi+2\left|\psi\right|^{2}\psi+\partial_{XX}\psi+\gamma\overline{\psi},\label{eq:pd-NLS}
\end{equation}
which is the parametric dissipative nonlinear Schrödinger equation
(pdNLSe). Notice that this equation is written in terms of the dimensionless
variables $T=\omega_{01}t$ and $X=b^{-1/2}kx$, where $t$ stands
for time, $x$ for the longitudinal spatial coordinate and $b=\frac{1}{4}\left[1+kd\left(1-\tau^{2}\right)/\tau\right]$.
The complex field $\psi\left(X,T\right)$ contains the slow spatiotemporal
modulation of the first transverse mode. In general, the deformation
at the free surface $\eta\left(x,y,t\right)$ and the velocity potential
inside the fluid, $\Phi\left(x,y,z,T\right)$, are related to $\psi\left(X,T\right)$
by means of
\begin{eqnarray}
\eta\left(x,y,t\right) & = & \Real\left\{ a\psi\left(X,T\right)\exp\mathrm{i}\omega t\right\} \sin ky,\label{eq:eta}\\
\Phi\left(x,y,z,t\right) & = & \Imag\left\{ a\psi\left(X,T\right)\exp\mathrm{i}\omega t\right\} \frac{g\sin ky\cosh k\left(z+d\right)}{\omega_{01}\cosh kd},\label{eq:Phi}
\end{eqnarray}
where $a^{2}=128k^{-2}/\left(6\tau^{2}-5+16\tau^{-2}-9\tau^{-4}\right)$.

Just as their propagating counterparts, non-propagating hydrodynamic
solitons arise from a subtle balance between linearities, nonlinearities
and dispersion. Parametrically excited solitary waves can be found
by assuming solutions of the form $\psi\left(X,T\right)=\rho\left(X\right)e^{-\mathrm{i}\theta}$.
Straightforward calculations show the onset of a subcritical instability
for $\gamma>\mu$ and $\nu<0$ with two families of solutions. One
family of solutions is always unstable whereas the other one,
\begin{equation}
\psi\left(X,T\right)=\pm\mathrm{i}\delta\sech\left[\delta\left(X-X_{0}\right)\right]e^{\frac{\mathrm{i}}{2}\sin^{-1}\frac{\mu}{\gamma}}\label{eq:soliton}
\end{equation}
is stable whenever $\gamma^{2}<\nu^{2}+\mu^{2}$; provided that $\delta^{2}=-\nu+\left(\gamma^{2}-\mu^{2}\right)^{\frac{1}{2}}$
\citep[see][]{1991JFM...223..589L}. The free parameter $x_{0}$ comes
out from a constant of integration and stands for the position of
the envelope centre of mass. Besides, (\ref{eq:soliton}) consists
of two solutions with opposite sign. This is consistent with experimental
observations of a soliton sort of \emph{polarity}, a crucial feature
for pair interactions \citep{1994PhLA..192....1W,1996PhLA..219...74W,2009RSPTA.367.3213C}.
Equation\ \ref{eq:pd-NLS} also supports cnoidal and dnoidal families
of solutions \citep[cf.][]{1984JFM...148..451M,1991JPSJ...60..146U}.
Besides, a change of sign in its nonlinear term gives rise to the
kink-type solutions observed by \citealp{1990PhRvL..64.1518D}. This
is achieved for instance by decreasing the depth of the fluid layer
\citep[see also][pp. 455-456]{1984JFM...148..451M}.

\section{Experimental setup\label{sec:Setup}}

\begin{figure}
\begin{centering}
\includegraphics{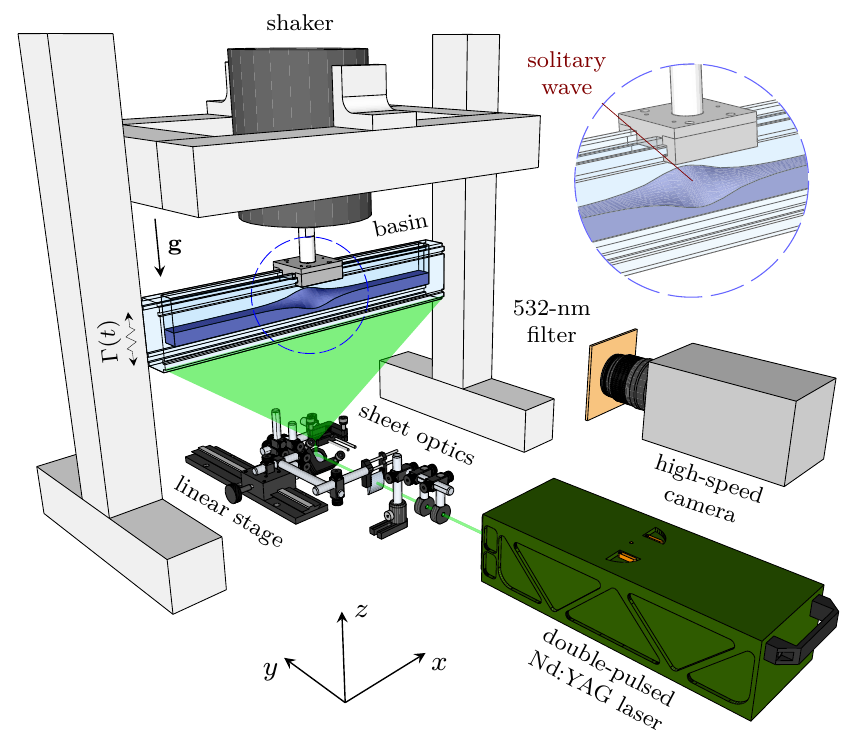}
\par\end{centering}

\caption{General scheme of the setup. Solitary waves are created in the fluid
contained by a basin subject to vertical vibrations. The fluid is
seeded with fluorescent particles and illuminated by a vertical laser
sheet shiftable in the $y$ direction. PIV digital processing provides
the velocity field in the $x$-$z$ plane at fixed $y_{s}$\label{fig:Setup}.}
\end{figure}

We run our experiments in an acrylic basin attached to an electromagnetic
shaker (see \figref{Setup}). The trough of length $l=19.05\;\mathrm{cm}$
and breadth $b=2.54\;\mathrm{cm}$ was filled with an aqueous solution
to a depth $d=2.00\;\mathrm{cm}$. The aqueous solution contained
$2\;\mathrm{ml}$ of Photoflo, used for improving wall wetting \citep{1984PhRvL..52.1421W},
and some KBr (13\% in mass concentration) for increasing the fluid
density. The solution density matches that of the PIV particles, $1.1\;\mathrm{g\cdot cm^{-3}}$,
so particle settling becomes noticeable only after several hours (kinematic
viscosity is also reduced in about 15\%). The acceleration of the
basin was registered using a piezoelectric accelerometer and a lock-in
amplifier, which was referenced externally to the input shaker signal.
Parametrically excited solitary waves can be observed when the system
is accelerated vertically as $\Gamma\left(t\right)=-\Gamma_{0}\cos\left(2\pi ft\right)$,
at frequencies $f$ slightly below $11\,\mathrm{Hz}$ and acceleration
amplitudes $\Gamma_{0}$ around $0.1g$. The frequency threshold is
very close to the the double of the experimental first transverse-mode
frequency, $f_{0,1}=5.49\,\mathrm{Hz}$, which was obtained by measuring
the linear surface response using a capacitive sensor and a spectrum
analyser in frequency-sweep mode (for more details, cf. \citet{Gordillo:2012wi}).

The PIV particles inside the fluid (carboxy-modified acrylate resin,
$\diameter=15\,\mu m$) were illuminated using a double-pulsed Nd:YAG
laser ($70\;\mbox{mJ}$ per pulse) and a laser sheet generator. We
placed the latter on a linear translational stage so the sheet position
along the fluid layer $y$ could be easily adjusted. The laser sheet
thickness inside the fluid was $2\,\mathrm{mm}$. Notice that illumination
from the bottom is the only one compatible with PIV and measurement
requirements. Unfortunately, due to the back and forth sloshing of
the solitary waves (see zoom window in \figref{Setup}), the free
surface reflects a huge amount of light on the $y$ direction. To
avoid this, we used fluorescent PIV particles (absorption and emission
peaks at $550$ and $580\,\mathrm{nm}$ respectively) and blocked
reflections with a longpass optical filter whose cut-off matches the
light-source wavelength ($532\,\mathrm{nm}$).

Images were acquired using a high-speed camera providing an imaging
region of $2560\times512\,\mathrm{pixels}$ ($20.0\times4.0\,\mathrm{cm^{2}}$).
We synchronized the laser double pulses $\left(\mbox{\ensuremath{\Delta}}t=10\;\mbox{ms}\right)$
with the motion of the solitary wave at some fixed phase $\theta_{s}$.
Since the solitary wave sloshes at $\frac{1}{2}f$ and at a fixed
phase with respect to the shaker input signal, the latter signal was
used as the reference. The frames were acquired synchronously with
the laser pulses. Each run consisted of 200 images pairs for fixed
sheet position $y_{s}$ and solitary-wave phase $\theta_{s}$. A whole
set of 36 different $\theta_{s}$ values were analysed throughout
the whole solitary-wave cycle. Besides, 10 laser sheet positions $y_{s}$
across the fluid layer were also analysed for a fixed phase, $\theta_{s}=\upi$,
at which the free-surface deformation is zero and the velocity is
maximal on the front wall.

Image processing was performed using our own Matlab code with classical
PIV digital techniques. Since image sequences included a moving boundary,
we required an automatic algorithm for boundary detection. We achieved
this by using a Radon-transform-based method from \citet{2011ExFl...51..871S}
on averaged same-phase samples. Likewise, background and illumination
issues were corrected using statistics-based images. The calculated
boundaries were then used for creating binary masks, which in turn
were used in a multi-pass interrogation PIV scheme. The minimum interrogation-window
size was $16\times16\;\mathrm{pixels}$$\left(1.25\times1.25\,\mathrm{mm^{2}}\right)$.
The results presented here were averaged over the 200 samples in the
correlation-function space to improve signal-to-noise ratio.

\clearpage{}

\section{Results\label{sec:Results}}

As a consequence of pulsed-laser synchronization, image sequences
are composed of a series of double fast snapshots $\left(\mbox{\ensuremath{\Delta}}t=10\;\mbox{ms}\right)$
captured every soliton period ($\mbox{\ensuremath{\Delta}}t'=2/f$).
This temporal scheme provides sets of time-resolved velocity fields
for a fixed phase between the external forcing and the soliton oscillation.
Tuning this phase, the velocity field during the whole cycle can be
found. The resulting velocity field $\left(u,v,w\right)$ is mainly
oscillatory in time, similar to those of stationary waves. But the
temporal scheme provides more than that. A simple inspection of sequences
shows that after one cycle, seeding particles do not return to their
position at the preceding cycle. If one frame is skipped such that
the temporal scheme is stroboscopic, it is easy to notice that particles
are constantly streamed. The effect of this streaming velocity field
$\left(\overline{u},\overline{v},\overline{w}\right)$ becomes perceptible
in particle trajectories only after one or several cycles. In this
sense, the instantaneous velocity field $\left(u,v,w\right)$ can
be considered as the sum of two components: the oscillatory part,
$\left(u-\overline{u},v-\overline{v},w-\overline{w}\right)$, and
the streaming one, $\left(\overline{u},\overline{v},\overline{w}\right)$.

\begin{figure}
\begin{centering}
\includegraphics{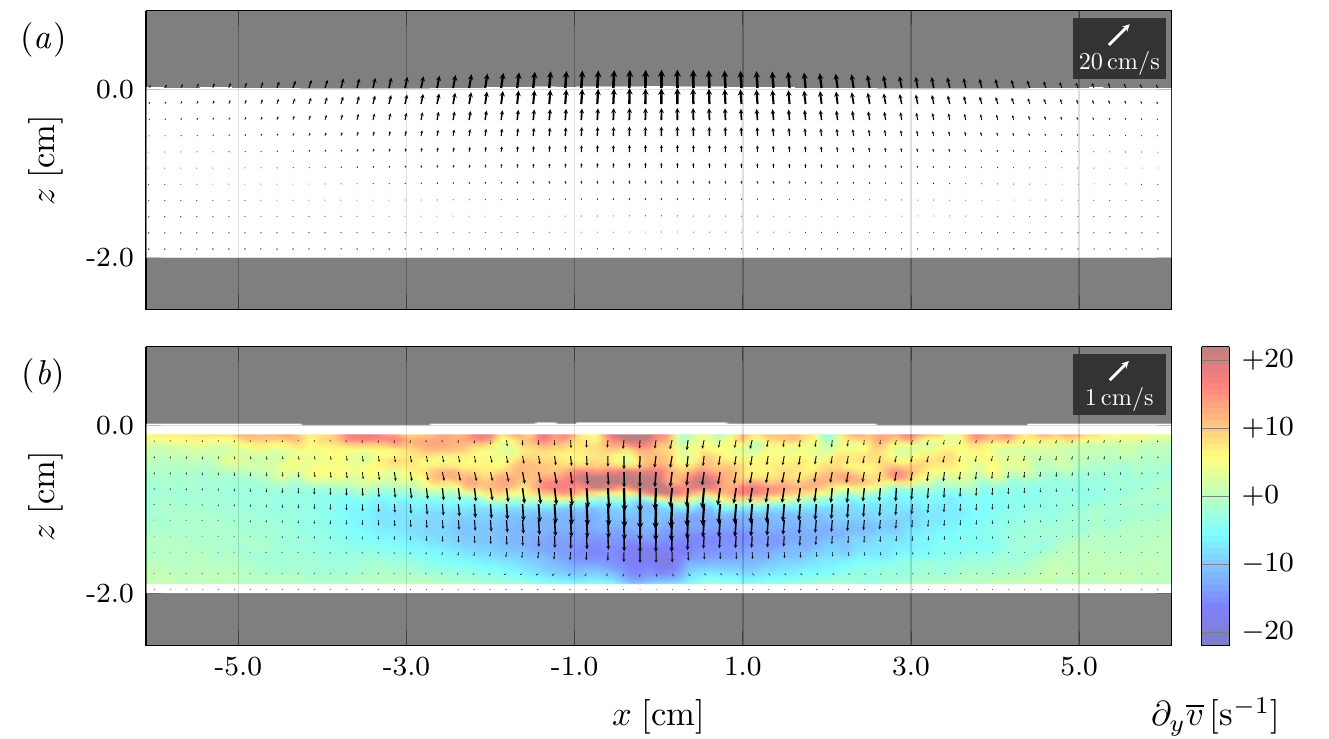}
\par\end{centering}

\caption{Front view ($x$-$y$ plane) of the velocity field of solitary wave
whose envelope is centred at $x=0$ and $\theta_{s}=\upi$. Only the
central region of the trough is shown. The laser sheet is placed at
$y_{s}=1.07\,\mathrm{cm}$. (\textit{a}) Instantaneous velocity field.
(\textit{b}) Streaming velocity field and out-of-plane velocity gradient
$\partial_{y}\overline{v}$ (in colours). The arrow scales for each
figure are also displayed. \label{fig:Field}}
\end{figure}

In figure\ \ref{fig:Field}, we depict the velocity field inside
the bulk of parametrically excited solitary wave at the phase $\theta_{s}=\upi$.
Figures\ \ref{fig:Field}(\textit{a,b}) display respectively the
instantaneous and the streaming velocity fields in the \emph{x-z}
plane for a fixed $y_{s}=-1.07\,\mathrm{cm}$. The driving frequency
and amplitude for the displayed solitary waves were $f=10.9\,\mathrm{Hz}$
and $\Gamma_{0}=0.096g$. The solitary-wave envelope is centred in
the basin ($x_{0}=0$ in equation \ref{eq:soliton}). Corresponding
videos can be viewed as supplementary material available online at:
\texttt{insert link}.

\begin{figure}
\begin{centering}
\includegraphics{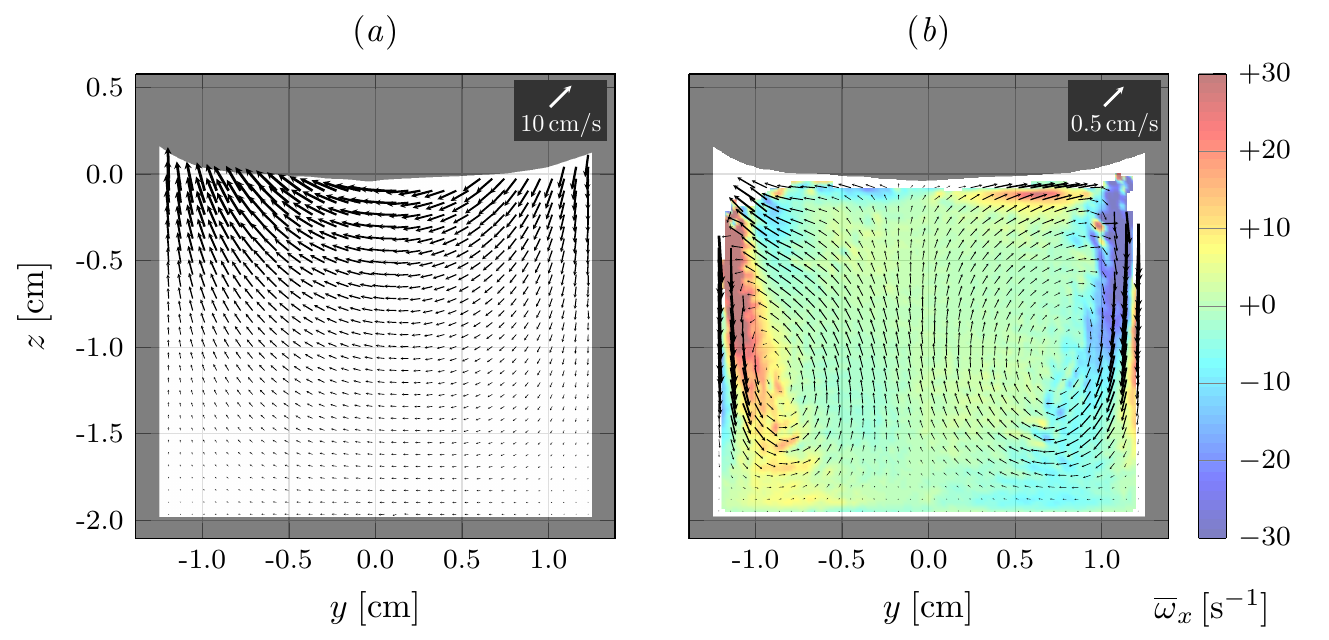}
\par\end{centering}

\caption{Front view ($x$-$z$ plane) of the velocity field of half solitary
wave pinned at a lateral wall at $\theta_{s}=\upi$. The laser sheet
was placed at $x_{s}=9.23\,\mathrm{mm}$. (\textit{a}) Instantaneous
velocity field. (\textit{b}) Streaming velocity field and out-of-plane
vorticity $\overline{\omega}_{x}$(in colours). \label{fig:Field-transverse}}
\end{figure}

In figure\ \ref{fig:Field-transverse}, we include also side views
($y$-$z$ plane) of the velocity field captured at the same phase
$\theta_{s}=\upi$. For this sequence, the basin was rotated 90º in
the $x-y$ plane (the axes remain fixed to the basin). To obtain a
full view of the velocity field of this plane, we used a half solitary
wave pinned at $x_{0}=\frac{1}{2}l$ instead of a centred one. In
this configuration and due to the refraction of rays of light on the
free surface, the uppermost region of a solitary wave centred at $x_{0}=0$
cannot be observed in images. Recalling that half solitary waves exist
in a particular region of parameters, the frequency and amplitude
of the external driving were suitably adjusted to \foreignlanguage{english}{$f=10.97\,\mathrm{Hz}$}
and $\Gamma_{0}=0.127g$. Figures\ \ref{fig:Field-transverse}(\textit{a,b})
display the instantaneous and the streaming velocity fields. For both
fields, the position of the laser sheet was $x_{s}=9.23\,\mathrm{cm}$.
On this side view, the zoomed images ($1280\times1600\,\mathrm{pixels}$
in a $2.8\times3.5\,\mathrm{cm^{2}}$ window; minimum interrogation
window: $16\times16\;\mathrm{pixels}$, \emph{i.e.} $0.35\times0.35\,\mathrm{mm^{2}}$)
allows to resolve the meniscus formed at the front and back walls.
In the figure, the white background represents image regions occupied
by the fluid. More videos are available online at: \texttt{insert
link}.


\subsection{Instantaneous velocity field}

The instantaneous velocity field inside the bulk of the parametrically
excited solitary wave is mainly given by an oscillatory part. Before
analysing any data, the uniform velocity due to the driving of the
basin was subtracted so the velocity fields are in a frame of reference
fixed to the basin. The front view of the velocity field shows that
the motion is highly localized in the envelope of the structure with
magnitudes decreasing one order of magnitude from the centre to the
side walls of the trough. The magnitude of the velocity increases
also as we approach the free surface, which is a general feature of
gravity waves in uniform depth containers. These two spatial features
can be observed through all the cycle of the solitary wave: the direction
of the arrows are mainly the same and only their magnitudes oscillate
in time. Thus, a single-phase snapshot, \emph{\eg }figure\ \ref{fig:Field}(\textit{a}),
provides a good overview of the distribution of velocity in the $x$-$y$
plane. Besides, a fast inspection of the instantaneous velocity field
at other planes by means of moving the laser sheet, shows that the
spatial and temporal features in the $x$-$z$ plane are the same.
The module of arrows just reduces as $y_{s}$ approaches the centre
of the basin in the $y$ direction.

The last feature is in agreement with the velocity fields obtained
from side views of the trough ($y$-$z$ plane) as depicted in figure\ \ref{fig:Field-transverse}(\textit{a}).
The temporal features of this view match those from the $x$-$z$
plane: a field with static orientation and time-oscillating magnitudes.
The snapshots also display that particles move from the positive side
of the $y$ axis to the negative one as the free surface acquires
its characteristic sloshing motion. The magnitudes increase when approaching
the free surface as expected.

\subsection{Streaming flow}

In contrast to the instantaneous velocity field, the streaming flow
does not oscillate during a cycle. By changing the phase of light
pulses with respect to the solitary-wave cycle, we observed that the
field is mainly steady across the bulk of the fluid. An example of
the velocity field in the $x$-$z$ is shown in figure~\ref{fig:Field}(\textit{b}).
The phase and $y_{s}$ position of the illuminated plane is the same
as for figure~\ref{fig:Field}(\textit{a}). Notice that compared
to the maximal instantaneous velocity, the magnitudes are around 20
times smaller. At this phase, we can observe that the particles are
pushed downward and outward the solitary-wave core. Higher magnitudes
are observed in the centre of the channel rather than close to the
free surface. The distribution of the velocity field suggests an important
out-of-plane velocity gradient(see colours in figure~\ref{fig:Field}\textit{b}).
Particles are streamed into the plane at the top and out of the plane
at the bottom. The opposite occurs at the centre of the basin ($y_{s}=0$),
where particles move upward everywhere.

To clarify the general structure of the streaming motion of particles,
the $y$-$z$ view is very useful. The velocity field in figure~\ref{fig:Field-transverse}(\textit{b})
shows a pair of vortex-like structures aligned to the $x$ axis. As
observed in the $x$-$z$ view, particles are streamed downward in
the front and back walls ($y=\pm\frac{1}{2}b$) and upward in the
centre of the trough. Streaming near both menisci is hard to resolve
since particle images are subject to high shear in this region. The
vorticity distribution (shown in colours in figure~\ref{fig:Field-transverse}\textit{b})
is highly localized near the front and back walls. The vorticity core
is pinched to the meniscus and slightly pushed back by the walls as
$z$ decreases.

\section{Discussion and conclusions\label{sec:Discussion}}

\subsection{Comparison with predicted results}

In general terms, the model of \citet{1984JFM...148..451M} based
on Hamiltonian equations and linear dissipation predicts well the
deformation of the free surface of parametrically excited solitary
waves \citep[see ][]{2009RSPTA.367.3213C,2011EPJD...62...39G}. Herein,
we display comparisons that now comprehend the velocity field inside
the bulk of solitary waves. According to \citeauthor{1984JFM...148..451M}
theory, the velocity of the fluid underneath the free surface is irrotational
and satisfies $\left(u,v,w\right)=\bnabla\Phi$. The potential is
given by
\begin{eqnarray}
\Phi\left(x,y,z,t\right) & = & \frac{\pm a\delta g\sin ky\cosh k\left(z+d\right)\cos\theta_{s}\left(t\right)}{\omega_{01}\cosh kd\cosh\left[\delta b^{-1/2}k\left(x-x_{0}\right)\right]},\label{eq:Phi-soliton}
\end{eqnarray}
where $\theta_{s}\left(t\right)=\omega t+\frac{1}{2}\sin^{-1}\frac{\mu}{\gamma}$.
Equation~(\ref{eq:Phi-soliton}) can be used to derive the predicted
components of the velocity. A simple way to test the accuracy of the
model is to fit velocity-field projections from model~(\ref{eq:Phi-soliton})
to the experimental data. To reduce the high dimensionality of the
set of dependant variables (three for space and one for time), we
fixed two dependant variables and applied a surface fit using the
\texttt{cftool} function in Matlab on the remaining two. 

\begin{figure}
\begin{centering}
\includegraphics{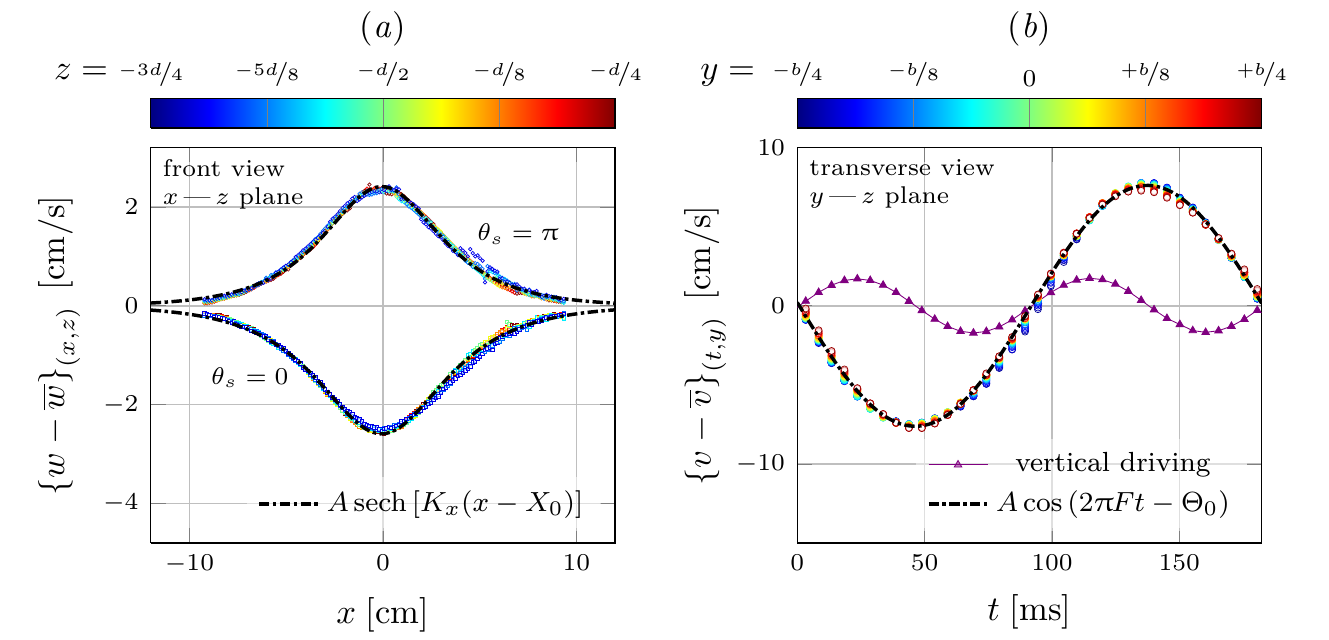}
\par\end{centering}

\caption{Surface fitting results of the potential model (\ref{eq:Phi-soliton})
for solitary waves. For visualization, surfaces are collapsed into
single master curves (symbols) that are then compared to fitted model
(dash-dotted lines). (\textit{a}) Maximal and minimal oscillatory
vertical velocity $w-\overline{w}$ in the$\left(x,z\right)$ plane.
Fitting parameters: (upper curve) $A=2.41\pm0.04\,\mathrm{cm/s}$,
$K_{x}=0.373\pm0.002\,\mathrm{cm^{-1}}$, $K_{z}=1.36\pm0.01\,\mathrm{cm^{-1}}$,
$X_{0}=-0.06\pm0.01\,\mathrm{cm}$; (lower curve) $A=-2.59\pm0.04\,\mathrm{cm/s}$,
$K_{x}=0.344\pm0.001\,\mathrm{cm^{-1}}$, $K_{z}=1.24\pm0.01\,\mathrm{cm^{-1}}$,
$X_{0}=-0.08\pm0.01\,\mathrm{cm}$, (\textit{b}) Oscillatory horizontal
velocity $v-\overline{v}$ in the $\left(y,t\right)$ plane. Fitting
parameters: $A=7.61\pm0.02\,\mathrm{cm/s}$, $f=5.492\pm0.005\,\mathrm{Hz}$,
$K_{y}=1.21\pm0.01\,\mathrm{cm^{-1}}$, $\Phi_{0}=-1.543\pm0.004$.
The basin vertical velocity is also displayed.\label{fig:fit}}
\end{figure}

In figure~\ref{fig:fit}(\textit{a}), we display the results for
the oscillatory vertical velocity $\left(w-\overline{w}\right)$ in
the $x$-$z$ plane. Here, $y$ is fixed at $y_{s}=1.07\,\mathrm{cm}$
and $t$ correspond to the two phases at which the velocities are
maximal and minimal, \emph{i.e. }$\theta_{s}=\left\{ 0,\upi\right\} $.
For visualization, we collapsed the PIV experimental data for eight
different $z$ values ($-1.5<z<0.5\,\mathrm{cm}$) into two master
curves. This was done by dividing the data by the $z$-dependant part
of $\partial_{z}\Phi$, \emph{i.e.} $\sinh\left[K_{z}\left(z+d\right)\right]$.
The curves should then be represented by a function $f\left(x\right)=A\sech K_{x}\left(x-X_{0}\right)$.
Variables in capital letters are fitted parameters. 

A similar comparison for the temporal evolution of the velocity field
of a half parametrically excited solitary wave is displayed in figure~\ref{fig:fit}(\textit{b}).
In this case, we analysed the oscillatory horizontal velocity from
the side views, $\left(v-\overline{v}\right)$, in terms of the time
$t$ and the horizontal coordinate $y$ at a fixed depth $z=-0.44\,\mathrm{cm}$
(the $x$ position is again fixed at $x_{s}=9.23\,\mathrm{mm}$).
For visualization, the surface in the $\left(y,t\right)$ space was
collapsed into a single master curve by diving the velocity profiles
by the $y$-dependence part $\partial_{y}\Phi,$\emph{ i.e.} $\cos\left[K_{y}y\right]$.
A set of 80 curves in the range $-0.64<y<0.64\,\mathrm{cm}$ were
used for this purpose. According to (\ref{eq:Phi-soliton}), the velocity
should be well fitted by $f\left(x\right)=A\cos\left(2\upi Ft-\Theta_{0}\right)$.
The vertical velocity of the basin is also plotted displaying the
parametric nature of the instability.

Figures \ref{fig:fit}(\textit{a-b}) are strong proofs that the potential-velocity
model describes well the oscillatory part of the velocity field. The
hyperbolic secant profiles reproduce with excellent accuracy the velocity
distribution along the solitary wave. The fitted values for maximal
and minimal vertical velocity $\left(w-\overline{w}\right)$ of \ref{fig:fit}(\textit{a})
show good agreement between them. In general, the fitted values for
$K_{y}$ and $K_{z}$ match well the crosswise wavenumber $k=\upi/b=1.24\,\mathrm{cm^{-1}}.$
Accordingly, experimental measurements for the half solitary wave
show a crosswise standing wave profile that oscillates harmonically
at the double of the driving period.

\subsection{Parametric streaming}

At this point, \citeauthor{1984JFM...148..451M}' model seems to be
very accurate to reproduce experimental data. At least, when comparisons
with experimental data are made after subtracting the mean streaming
velocity field to the instantaneous one. This important step of processing
is vital for the good agreement displayed in figures \ref{fig:fit}(\textit{a-b}).
Although \citeauthor{1984JFM...148..451M}' model for Hamiltonian
flows can yield some sort of streaming for higher-order terms \citep[see][]{Gordillo:2012wi},
these corrections remain potential across the whole bulk of the fluid.
Hence, the model is blind to streaming flows with a vorticity distribution
as the one that parametrically excited solitary waves support. Even
more striking is the fact that the streaming flow is so significant.
Using equation (\ref{eq:Phi}) and the formulas for the pdNLSe parameters,
we can estimate roughly the following values from experimental data:
$\psi\sim10^{-\text{1}}$ and $\nu\sim\mu\sim\gamma\sim10^{-\text{1}}$.
Meanwhile, the ratio between the streaming and the potential velocities
is $\overline{w}/w\approx10^{-1}$. This means that corrections due
to the coupling of streaming and potential-flow are $\psi^{3}$ and
hence, as important as the the higher-order term in (\ref{eq:pd-NLS}),
which is actually responsible for the highly localized envelope of
parametrically excited solitary waves. Two question naturally arises:
What is the physical origin of the parametric streaming? Why Miles
model and pdNLSe have been so successful in describing parametrically
excited solitary waves despite they do not consider the streaming
observed in experiments?

The physical origin of streaming in parametric flows has been widely
studied in a domain slightly different: acoustic flows. Streaming
occurs because oscillatory boundary layers transfer vorticity to the
bulk of the fluid \citep[see][pp. 358-361 and references therein]{Batchelor2000an_introduction}.
The induced streaming velocity field is independent of the viscosity
of the fluid $\mu_{0}$ and does not vanish as $\mu_{0}\rightarrow0$
as a consequence of the singular limit of the Navier-Stokes equation
in the high Reynolds number limit. Streaming in parametric instabilities
were first visualized by \citet{1990JFM...221..383D} using Kalliroscope
particles in a Faraday-instability configuration. To our knowledge,
quantitative measurements of the streaming field in a parametric instability
have never been reported. Theoretical analysis on this subject are
also rare and have addressed to the bottom and free-surface boundary
layers \citep{2002JFM...467...57M,2005JFM...546..203M}. Since parametrically
excited solitary waves are necessarily supported between two vertical
walls, we are sceptical about their applicability to our setup. Furthermore,
it is clear that the effect of the advancing and receding menisci
cannot be disregarded as they are crucial for the correct characterization
of the vorticity field (see figure~\ref{fig:Field-transverse}\emph{b}). 

Concerning the second question, the few theoretical works made on
this subject can give us important clues. \citet{2002JFM...467...57M}
considered the problem of finding an amplitude equation for two-dimensional
Faraday waves starting from Navier-Stokes equation in a laterally
unbounded fluid. The resulting amplitude equation is similar to that
obtained from Hamiltonian formulations \citet{1984JFM...146..285M}
except for an integral term that accounts for the coupling between
streaming and the Faraday waves. This term does not generate a major
change in the general dynamics of the amplitude equation although
its vital to explain drift instability in Faraday waves \citep{2002JFM...467...57M}.
For parametrically excited solitary waves, although calculations are
considerably more complex, streaming coupling should provide corrections
to the $a$ factor, defined after equations (\ref{eq:eta}) and (\ref{eq:Phi}),
without modifying the dynamics of the pdNLSe equation. However, it
is hard to establish the scope of this coupling term for further bifurcations.
In order to accomplish the challenge of incorporating parametric streaming
in the amplitude equation for parametrically excited solitary waves,
we should undoubtedly address to a more fundamental hydrodynamical
problem: understand and be able to predict streaming near oscillatory
contact lines.

\section*{Acknowledgements}

We are thankful to Edgar Knobloch and Marcel Clerc for fruitful discussions.
The research was supported by Conicyt grants ACT 127 and AIC 43. LG
acknowledges Conicyt fellowships 57080094 and 24100131, and the AXA
Research Fund.

\section*{Supplementary movies}

Supplementary movies are available at ...

\bibliographystyle{jfm}
\bibliography{BibTex/new_books,BibTex/papers20012013}

\end{document}